\documentclass[aps,prl,twocolumn,showpacs,superscriptaddress]{revtex4}
\usepackage{epsfig}
\usepackage{graphicx}
\usepackage{amsfonts}
\usepackage[figuresright]{rotating}
\usepackage{amssymb}
\usepackage{amsmath}
\usepackage{dcolumn}
\usepackage{bm}

\def\be{\begin{equation}} \def\ee{\end{equation}}
\def\bea{\begin{eqnarray}} \def\eea{\end{eqnarray}}

\def\nn{\nonumber}
\def\pp{\parallel}

\begin{document}
\title{Stable Fulde-Ferrell-Larkin-Ovchinnikov pairing states in 2D
and 3D optical lattices}
\author{Zi Cai}
\affiliation{Department of Physics, University of California,
San Diego, CA92093}
\author{Yupeng Wang}
\affiliation{Beijing National Laboratory for Condensed Matter
Physics, Institute of Physics, Chinese Academy of Sciences, Beijing
100080, P. R. China}
\author{Congjun Wu}
\affiliation{Department of Physics, University of California,
San Diego, CA92093}
\begin{abstract}
We present the study of the Fulde-Ferrell-Larkin-Ovchinnikov (FFLO) pairing
states in the $p$-orbital bands in both two and three-dimensional
optical lattices.
Due to the quasi one-dimensional band structure which arises from the
unidirectional hopping of the orthogonal $p$-orbitals, the pairing phase
space is not affected by spin imbalance.
Furthermore, interactions build up high dimensional phase coherence which
stabilizes the FFLO states in 2D and 3D optical lattices in a large
parameter regime in the phase diagram.
These FFLO phases are stable with imposing the inhomogeneous trapping potential.
Their entropies are comparable to the normal states at finite
temperatures.
\end{abstract}
\maketitle

The FFLO phases are a class of exotic Cooper pairing states exhibiting
non-zero center of mass momenta \cite{fulde1964,larkin1965,casalbuoni2004,
matsuda2007}, which occur in spin imbalanced systems with mismatched
Fermi surfaces.
However, such states are difficult to realize in solid state systems.
The strong orbital effects of external magnetic fields often suppress
Cooper pairing before sizable spin polarizations are reached.
Moreover, because only small fractions of the mismatched Fermi
surfaces can participate pairing, the FFLO states are usually
fragile in 2D and 3D systems.
In spite of indirect evidence in various heavy fermion compounds and
organic superconductors (e.g. CeCoIn$_5$
\cite{cecoin5} and $\lambda$-(BETS)$_2$FeCl$_4$ \cite{uji2001}),
the FFLO states remain elusive.

In the cold atom community, the search for the FFLO pairing states
has been attracting considerable interest \cite{liao2009,partridge2006,
ketterle2008,radzihovsky2009,loh2010,zhao2008a, koponen2007,
koponen2008, parish2007,feiguin2007,batrouni2008,edge2009,
LiuXJ2007,
paananen2008,luescher2008,kakashvili2009,yang2005,kinnunen2006,
mizushima2005,bakhtiari2008,nikolic2010}. Spin imbalanced
two-component fermion systems have been prepared free of
the orbital effects of magnetic fields. However, the
problem of the limited pairing phase space remains, thus phase
separations are observed experimentally instead of the FFLO pairing
in 3D traps \cite{zwierlein2006,partridge2006}. This difficulty is
avoided in 1D systems whose Fermi surfaces are points, thus spin
imbalance does not affect the pairing phase space. Considerable
progress has been made in quasi-1D systems of coupled optical tubes
in Hulet's group \cite{liao2009}, in which the partially polarized
central regions in the tubes are observed in agreement with the
prediction of the Bethe ansatz solution. However, due to the
intrinsic strong quantum fluctuations in 1D, the pairing density
waves, which are the smoking gun evidence for the FFLO states,
cannot be long-range-ordered and thus difficult to observe.

On the other hand, orbital physics with cold atoms in optical
lattices has received considerable attention, which gives rise to a
variety of new states of matter with both cold bosons and fermions
\cite{isacsson2005,liu2006,kuklov2006,stojanovic2008,
mueller2007,wirth2010}. In particular, it has been recently shown
that the $p_{x,y}$-orbital band in the honeycomb lattice exhibits
different properties from its $p_z$-orbital counterpart of graphene.
These include the strong correlation effects in the flat bands (e.g.
Wigner crystallization  \cite{wu2007} and ferromagnetism
\cite{zhang2008}), quantum anomalous Hall states
\cite{zhangmachi2010}, and the heavily frustrated orbital
exchange physics \cite{wu2008,chern2011}.

In this article, we combine the realization of the FFLO states
and the study of orbital physics with cold atoms together.
The FFLO states can be stabilized in the $p$-orbital
bands in both 2D square and 3D cubic optical lattices.
Different from the metastable $p$-orbital boson systems
\cite{mueller2007,wirth2010}, the $p$-orbital systems filled with
fermions with the fully filled $s$-band are stable due to Pauli's exclusion
principle.
This work is a natural high dimensional generalization of the current
experiments in Hulet's group \cite{liao2009}.
The $p_{x}~(p_y, p_z)$-orbital bands behave like orthogonally-crossed
quasi-1D arrays due to their highly unidirectional hoppings.
The onsite negative Hubbard interactions further build up high dimensional
phase coherence over the entire lattice.
It combines the advantages of the large pairing phase space of
quasi-1D systems and the high dimensional phase coherence.

\begin{figure}[htb]
\includegraphics[width=0.8\linewidth]{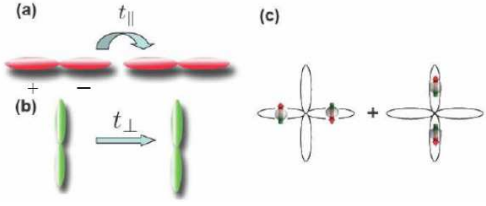}
\caption{(Color online) (a) and (b) describe the longitudinal
hopping $t_\parallel$-term and the transverse $t_\perp$-term  of the
the $p$-orbitals, respectively. (c) The pairing hopping term in Eq.
(\ref{eq:porbital_int}) locks the phases of two onsite intra-orbital
pairings in the $p_x$ and $p_y$-orbitals. } \label{fig:FFLO}
\end{figure}

The anisotropic $p$-orbital bands possess the quasi-1D like structures
with perfect nesting at general fillings and spin imbalance.
For simplicity, we start with the 2D case.
The similar physics applies to the 3D cubic lattice as well.
We present the $p$-band Hamiltonian as
\bea
H_0 &=& t_\parallel \sum_{\vec r,\alpha}
\Big\{p^\dagger_{x,\alpha}(\vec r) p_{x,\alpha} (\vec r+\hat e_x)
+ p^\dagger_{y,\alpha}(\vec r)
p_{y,\alpha}( \vec r+ \hat e_y) \Big\} \nn \\
&-&
\mu \sum_{\vec r,\alpha} n_{\alpha} (\vec r) -\frac{h}{2} \sum_{\vec r}
\big\{ n_\uparrow (\vec r) -n_\downarrow (\vec r) \big\},
\label{eq:bandsquare}
\eea
where $\alpha$ refers to spin index; $h$ controls spin imbalance;
$n_\alpha(\vec r)=p^\dagger_{x,\alpha}(\vec r) p_{x,\alpha}(\vec r)
+p^\dagger_{y,\alpha}(\vec r) p_{y,\alpha}(\vec r)$  is the particle
number of spin $\alpha$.
Only the longitudinal $\sigma$-bonding $(t_\parallel)$ term is kept which
describes the hopping between $p$-orbitals along the bond direction as
depicted in Fig. \ref{fig:FFLO} (a).
$t_\parallel$ is positive because of the odd parity of the $p$-orbitals.
The transverse $\pi$-bonding term with the hopping integral $t_\perp$
is neglected, which describes the hopping between $p$-orbitals
perpendicular to the bond direction as depicted in Fig. \ref{fig:FFLO}
(b).

\begin{figure}[htb]
\includegraphics[width=0.49\linewidth]{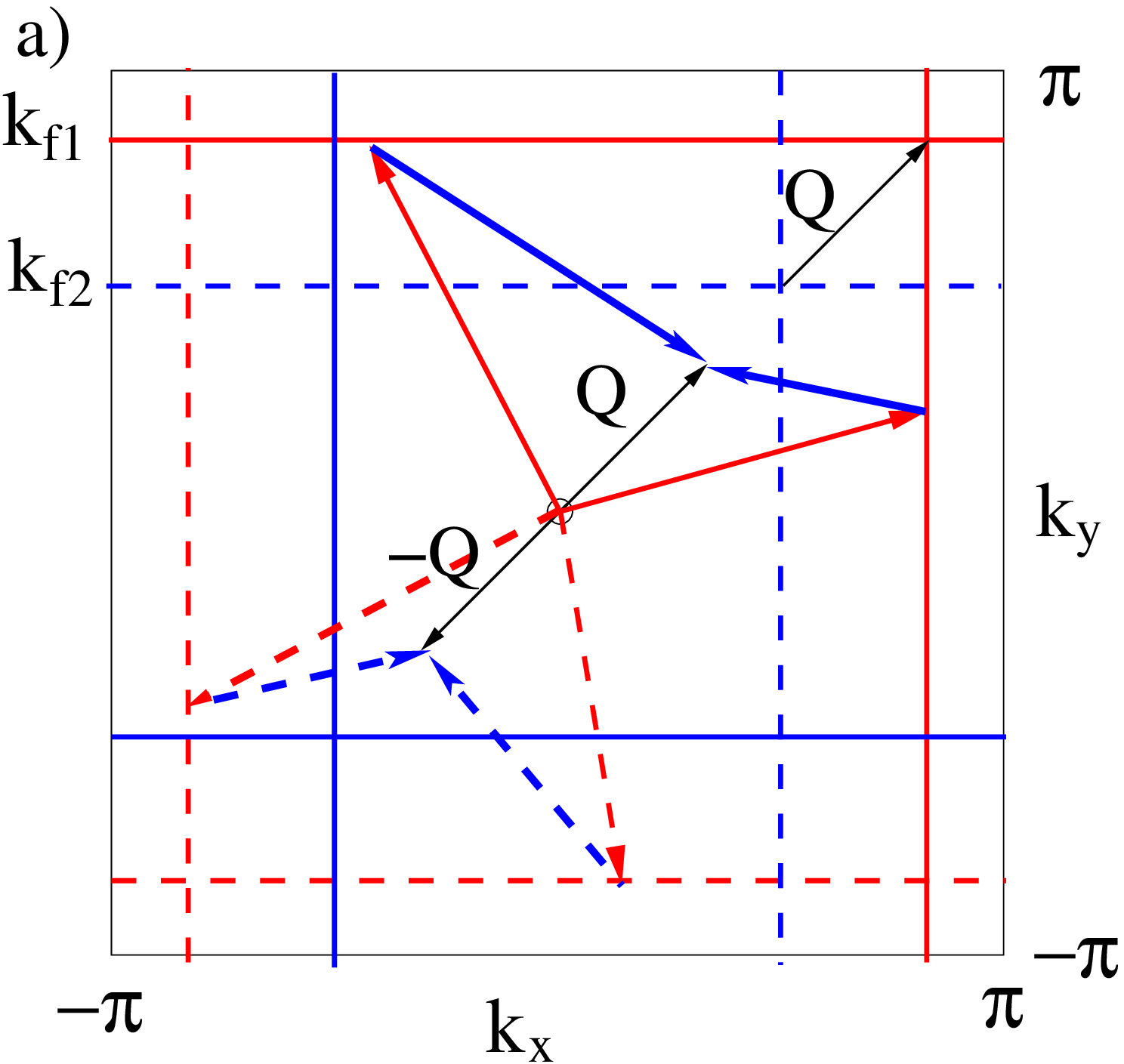}
\includegraphics[width=0.49\linewidth]{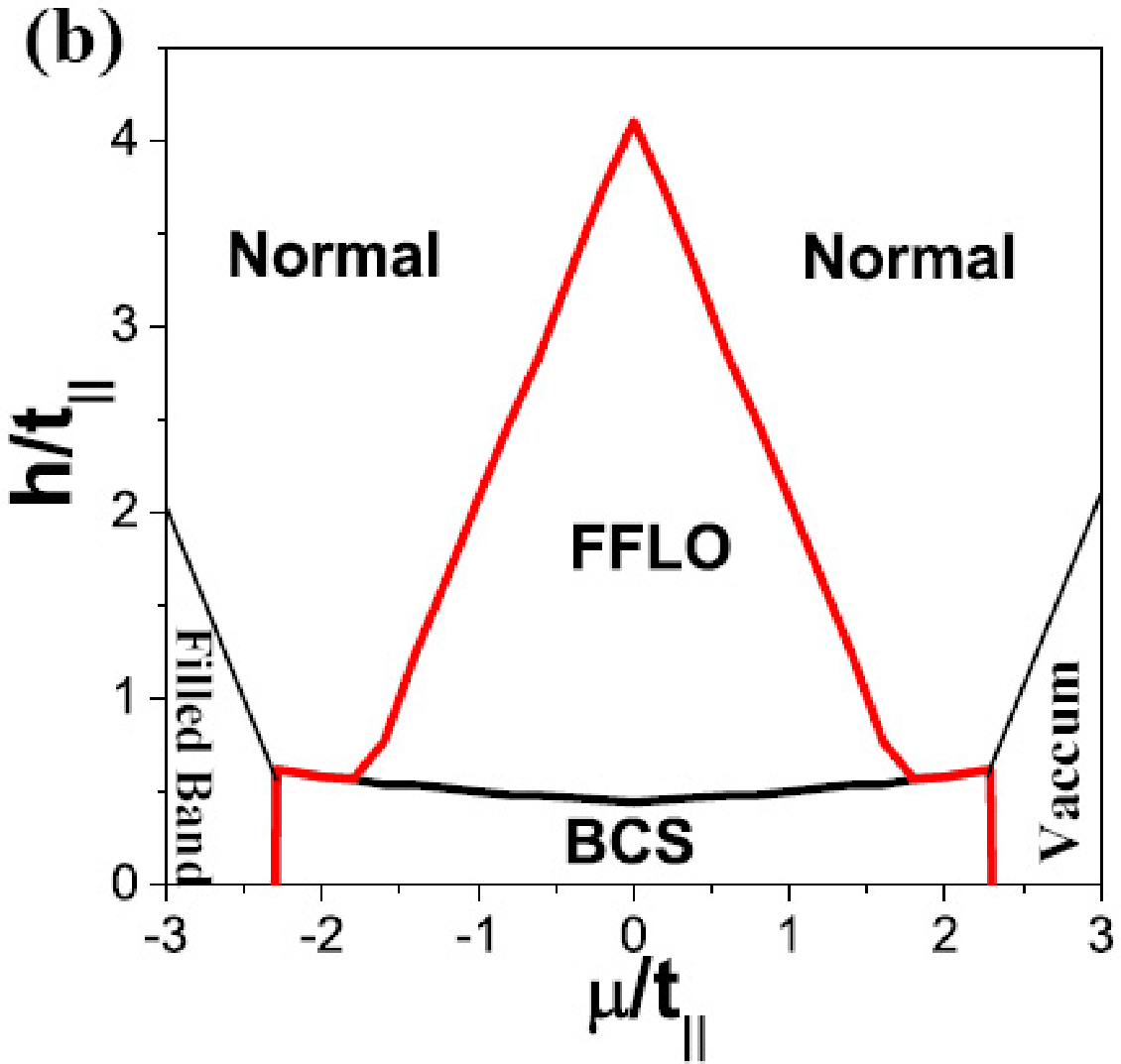}
\caption{(Color online) (a)The nesting of the $p$-orbital Fermi
surfaces ensures that all of the Fermi surfaces are paired at a
general filling and spin imbalance. The Fermi surfaces of the $p_x$
($p_y$)-orbitals are vertical (horizontal) lines; those of the
majority (minority) spins are marked red (blue). Fermi surfaces
marked with solid (dashed) lines are paired with the center of mass
momentum $\pm\vec Q$, respectively. The red and blue arrows
represents the Fermi wavevectors of spin up and down fermions
participating Cooper pairing. (b) The 2D phase diagram from the B-de
G solution as chemical potential $\mu$ and the magnetic field $h$
with $U/t_\parallel=-1.5$. } \label{fig:phase}
\end{figure}

In spite of the 2D lattice structure, the $p$-orbital band structure
of Eq. \ref{eq:bandsquare} remains quasi-1D-like as depicted in Fig.
\ref{fig:phase} (a).
The $p_x (p_y)$-orbital band disperses along the $x (y)$-direction,
respectively, but does not along the $y (x)$-direction.
The Fermi surfaces are vertical ($p_x$) and horizontal
$(p_y)$ lines across the entire Brillouin zone.
For the arbitrary filling and spin imbalance, the Fermi surfaces of
spin up and down fermions have the perfect nesting.
Consequentially, spin imbalance {\it does not} suppress the pairing
phase volume.
The high dimensional $p$-orbital systems
have the same advantage as that in 1D systems.

The important feature of the 2D $p$-orbital systems for the FFLO states
is that the onsite negative Hubbard interactions build up the 2D phase
coherence.
The interactions are represented in the standard two-orbital Hubbard
model as
\bea
H_{int}&=& \sum_{\vec r} U \Big [n_{ x\uparrow } (\vec r) n_{x\downarrow }
(\vec r) +n_{y\uparrow }(\vec r) n_{y\downarrow } (\vec r) \Big ] \nn \\
&-&\sum_{\vec r}J \Big [\vec S_{x}(\vec r) \cdot
\vec S_{y}(\vec r)
-\frac{1}{4} n_{x} (\vec r) n_{y} (\vec r) \Big] \nn \\
&+&\sum_{\vec r}\Delta \Big [p^\dagger_{x\uparrow}
(\vec r) p^\dagger_{x\downarrow} (\vec r) p_{y\downarrow} (\vec r)
p_{y\uparrow}(\vec r)+h.c. \Big ],
\label{eq:porbital_int}
\end{eqnarray}
where $U=g\int dr |\psi_{p_{x,y}}(\vec r)|^4<0$ and $g$ is the contact
interaction in the $s$-wave scattering approximation.
$J$ and $\Delta$ satisfy $J=\frac{2U}{3}<0$ and $\Delta=\frac{U}{3}<0$
\cite{zhang2008}.
The negative $U$-term gives rise the dominant intra-orbital singlet
pairings in the $p_x$ and $p_y$-orbitals, defined as
\bea
\Delta_x(\vec r)&=&\langle
G|p_{x\uparrow}(\vec r)p_{x\downarrow}(\vec r)|G\rangle, \nn \\
\Delta_y(\vec r)&=&\langle G|p_{y\uparrow}(\vec r)p_{y\downarrow}
(\vec r)|G\rangle,
\eea
where $|G\rangle$ is the mean field pairing ground states.
The $J$-term induces the inter-orbital singlet pairing between
$p_x$ and $p_y$-orbitals.
However, because the Fermi surfaces of $p_x$ and $p_y$-orbitals
are orthogonal, the inter-orbital pairing is unfavorable.

The pair hopping $\Delta$-term in Eq. \ref{eq:porbital_int} can
be considered as the internal Josephson coupling to lock
the phases of two intra-orbital pairings $\Delta_x$ and $\Delta_y$.
As a result, the motion of Cooper pairs are 2D-like in spite
of the quasi 1D-like single fermion hopping.
To clarify the pairing symmetry, we first consider two fermions on the
same site to gain some intuition.
The $s$-wave Feshbach resonances forbid spin triplet channel and  induce
a spin singlet pairing.
In the spin singlet channel, their orbital wavefunctions are symmetric
as $p_x^2+p_y^2$, $p_x^2-p_y^2$ and $p_xp_y$ respectively.
The first one has energy $U +\Delta = 4U/3$, while the later two are
degenerate with energy $U-\Delta=J =2U/3$.
From this simple analysis, we can see that the system favors pairing
with $p_x^2+p_y^2$ orbital symmetry (as shown in Fig. \ref{fig:FFLO}(c)),
while pairing with other two symmetries are suppressed, which can be
verified by the numerical results below.

We have performed calculations based on the self-consistent Bogoliubov-de
Gennes (B-deG) solution to study the competition among the FFLO state,
the BCS state and the normal state as presented in Fig. \ref{fig:phase} (b).
To synchronize the phases of $\Delta_x(\vec r)$ and $\Delta_y(\vec r)$
on each site, their center of mass wavevectors in the FFLO states
have to be the same.
This can be achieved by choosing the pair density wavevectors along
the diagonal direction $\pm\vec Q$ defined as $\vec Q=(\delta k_f,
\delta k_f)$ where $\delta k_f=k_{f_1}-k_{f_2}$, and $k_{f_{1,2}}$ are
Fermi wavevectors of the majority and minority spins as indicated
in Fig. \ref{fig:phase} (a).
By the symmetry of the square lattice, $\vec Q^\prime=\pm(\delta
k_f, -\delta k_f)$ are another possible choice of pair density
wavevector.
We consider the simplest Larkin-Ovchinnikov (LO) states with one
pair of Cooper pair momenta $\pm \vec Q$, with the
sinusoidal order parameter configuration as
\bea
\Delta_x(\vec r)=\Delta_y(\vec r) =|\Delta| \cos (\vec Q \cdot \vec r).
\label{eq:delta}
\eea
The LO state breaks both translational and the 4-fold lattice rotational
symmetries.
We have performed unbiased real space B-de G calculations without 
specifying the FFLO momentum in the initial conditions but rather
starting from a configuration with uniform pairing.
The FFLO momentum $Q$ in the above analysis is obtained when the
numerical convergence is arrived.
Compared with the phase diagram of spin-imbalanced fermions in
the $s$-orbital band, the FFLO phase in our $p$-orbital band system
exists in a much larger regime in the phase diagram sandwiched
between the fully paired BCS phase and the fully polarized normal phase.

\begin{figure}[htb]
\includegraphics[width=0.485\linewidth]{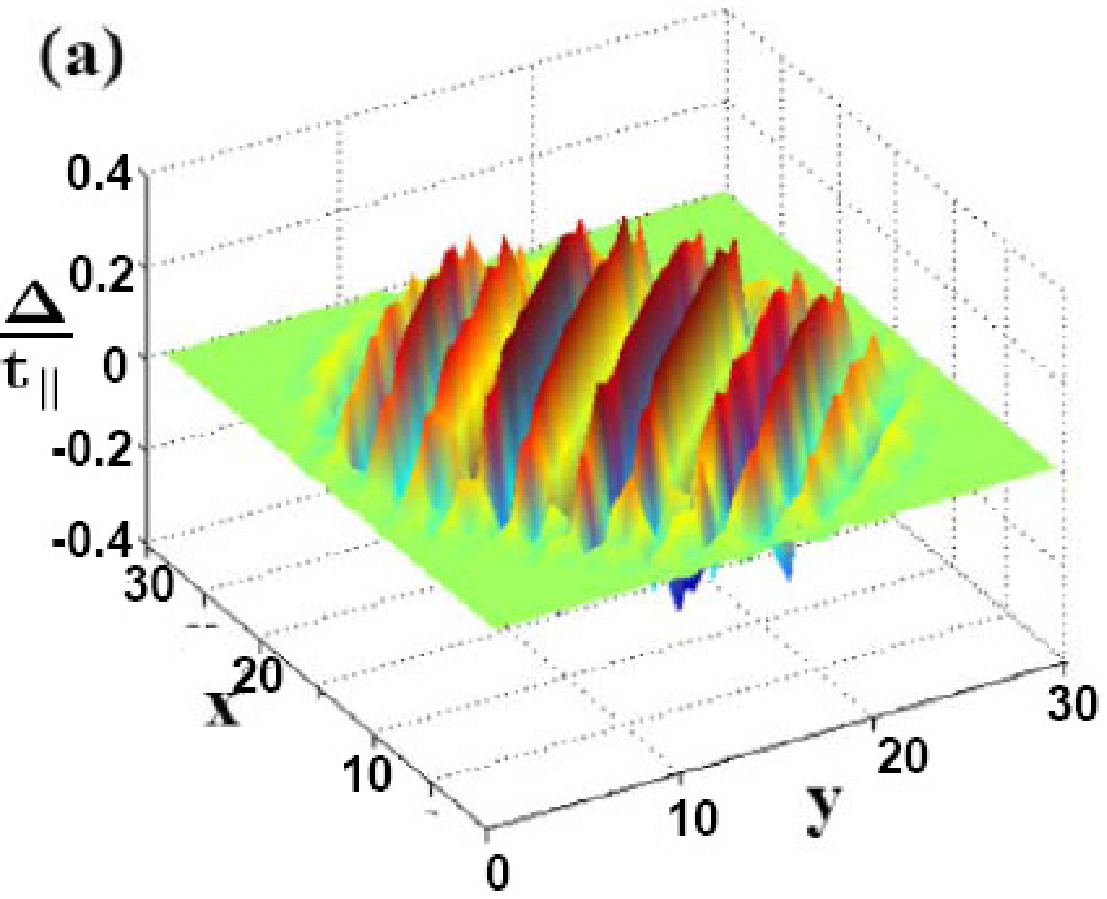}
\includegraphics[width=0.49\linewidth]{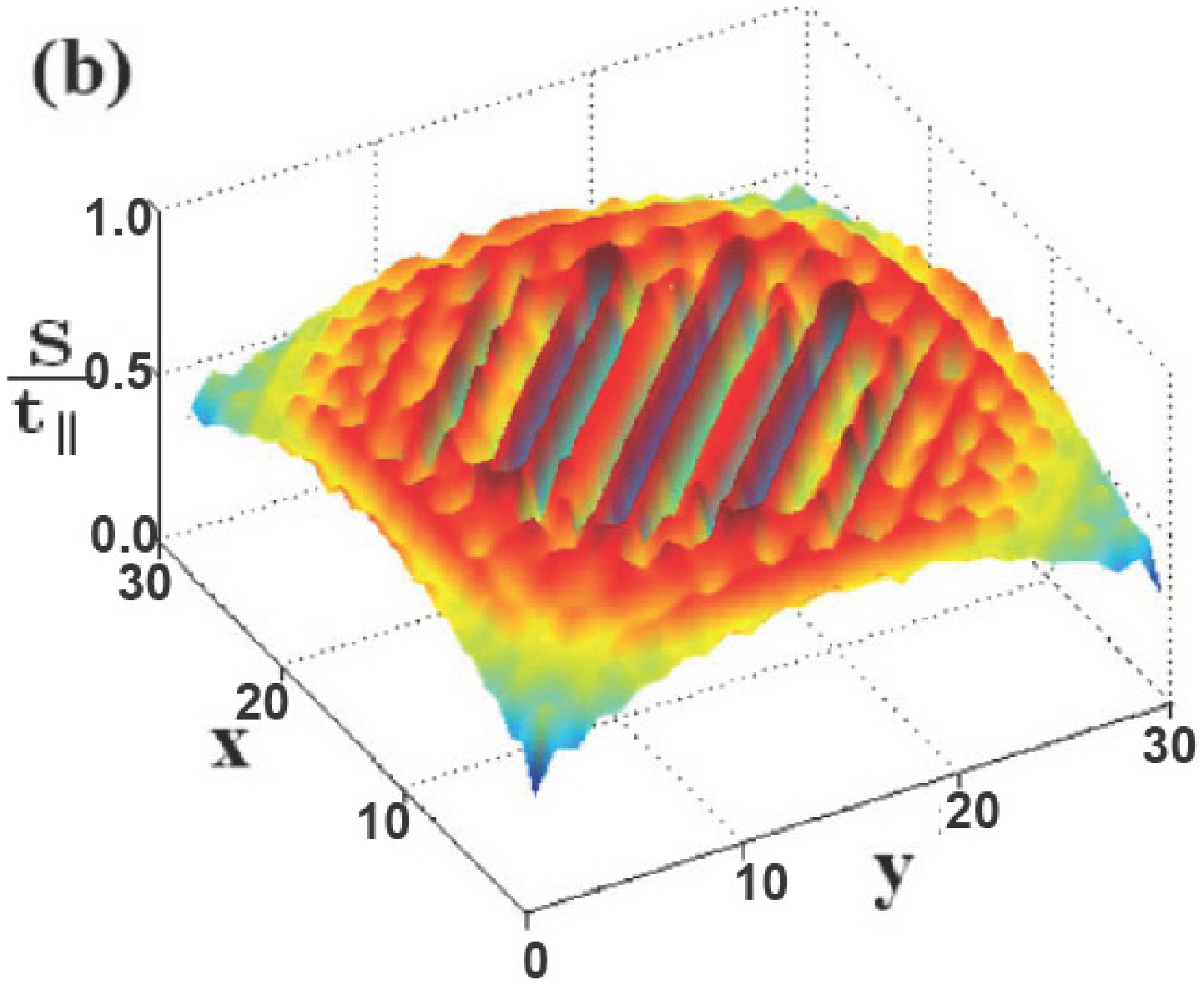}
\caption{(Color online) The B-de G solution of the $p$-orbital FFLO
state in a $30\times 30$ lattice with a weak confining trap. (a) The
order parameter distribution of $\Delta_x(\vec r)$ oscillates  along
the $[\bar{1}1]$-direction. (b) The spin density distribution
$S_z(\vec r)$ peaks around the gap nodes.} \label{fig:delta}
\end{figure}

Next we study a more realistic situation, the effects of the soft
confining potential to the $p$-orbital FFLO states, by performing
self-consistent real space B-de G calculations. We consider a
$30\times 30$ lattice with the harmonic trapping potential $V_{ex}=
A (r/a)^2$ where $A/t_\parallel=5\times 10^{-3}$; $r$ is the
distance from the trap center; $a$ is the lattice constant. The real
space distribution of order parameter $\Delta_x(\vec r)$ is shown in
Fig.\ref{fig:delta} (a) with the parameters chosen as
$h/t_\parallel=3$, $U/t_\parallel=-3$ and $\mu=0$. Clearly
$\Delta_x$ oscillates along the $[\bar{1}1]$-direction in agreement
with the previous analysis. To verify Eq. \ref{eq:delta}, we further
calculate the difference between the pairing orders in different
orbitals. In the bulk, the relation that $\Delta_x (\vec
r)=\Delta_y(\vec r)$ is well-satisfied. The difference between
$\Delta_x(\vec r)$ and $\Delta_y(\vec r)$ is only important at the
boundary which breaks the symmetry between the $p_x$ and
$p_y$-orbitals. The spin density distribution $s_z(\vec
r)=n_\uparrow(\vec r) -n_\downarrow (\vec r)$ is depicted in Fig.
\ref{fig:delta} (b). It peaks around the gap nodes, which is
consistent with the fact that spin polarization suppresses Cooper
pairing.

Next we discuss the effect of the small $\pi$-bonding $t_\perp$, which
has been neglected above but always exists in realistic systems.
The $t_\perp$-term restores the 2D nature of the Fermi surfaces
and suppresses the perfect nesting, therefore it is harmful to
the FFLO states.
Our numerical result indicates that the FFLO state remains
stable at small values of $t_\perp$.
For example, with $U/t_\pp=-3$, $h/t_\pp=3$ and $\mu=0$, the FFLO
state survives until $t_\perp/t_\parallel$ reaches $0.12$.
Beyond this value, it changes
to the normal state through a first order phase transition.
As calculated in Ref. \cite{isacsson2005}, with the optical potential depth
$V_0/E_R\approx 15$, $t_\parallel$ is at the order of $0.1 E_R$ and
$t_\perp/t_\parallel \approx 5\%$.
Increasing optical potential depth further suppresses $t_\perp$, thus there
is a large parameter regime to stabilize the FFLO states.


\begin{figure}[htb]
\includegraphics[width=0.8\linewidth]{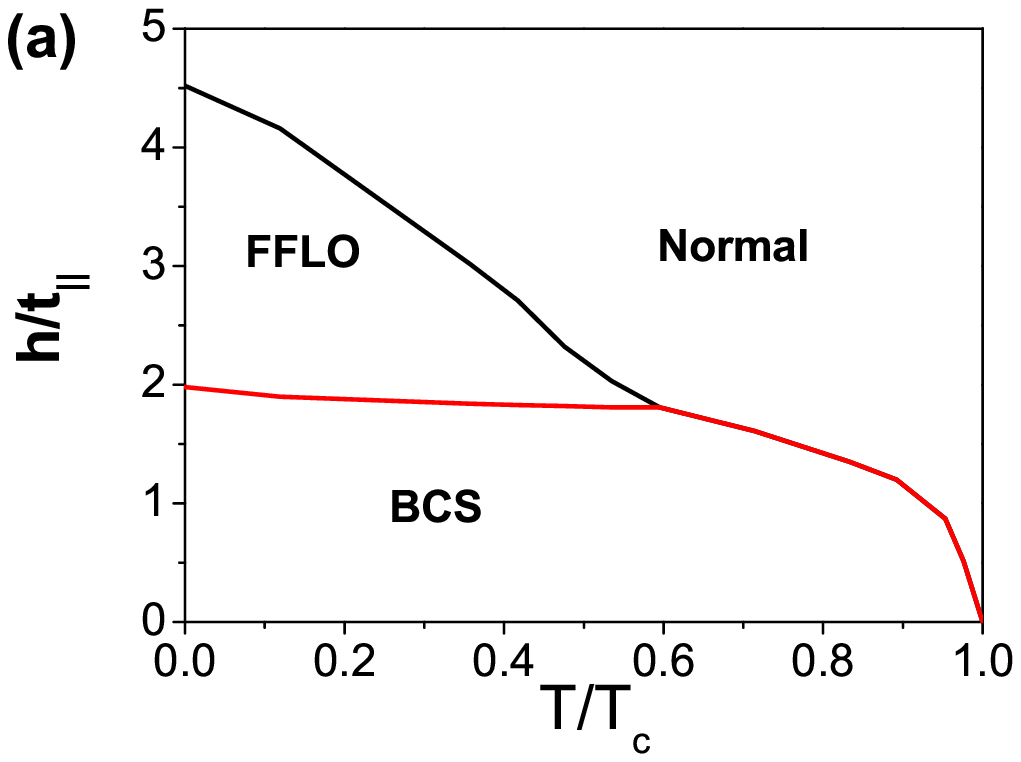}
\includegraphics[width=0.7\linewidth]{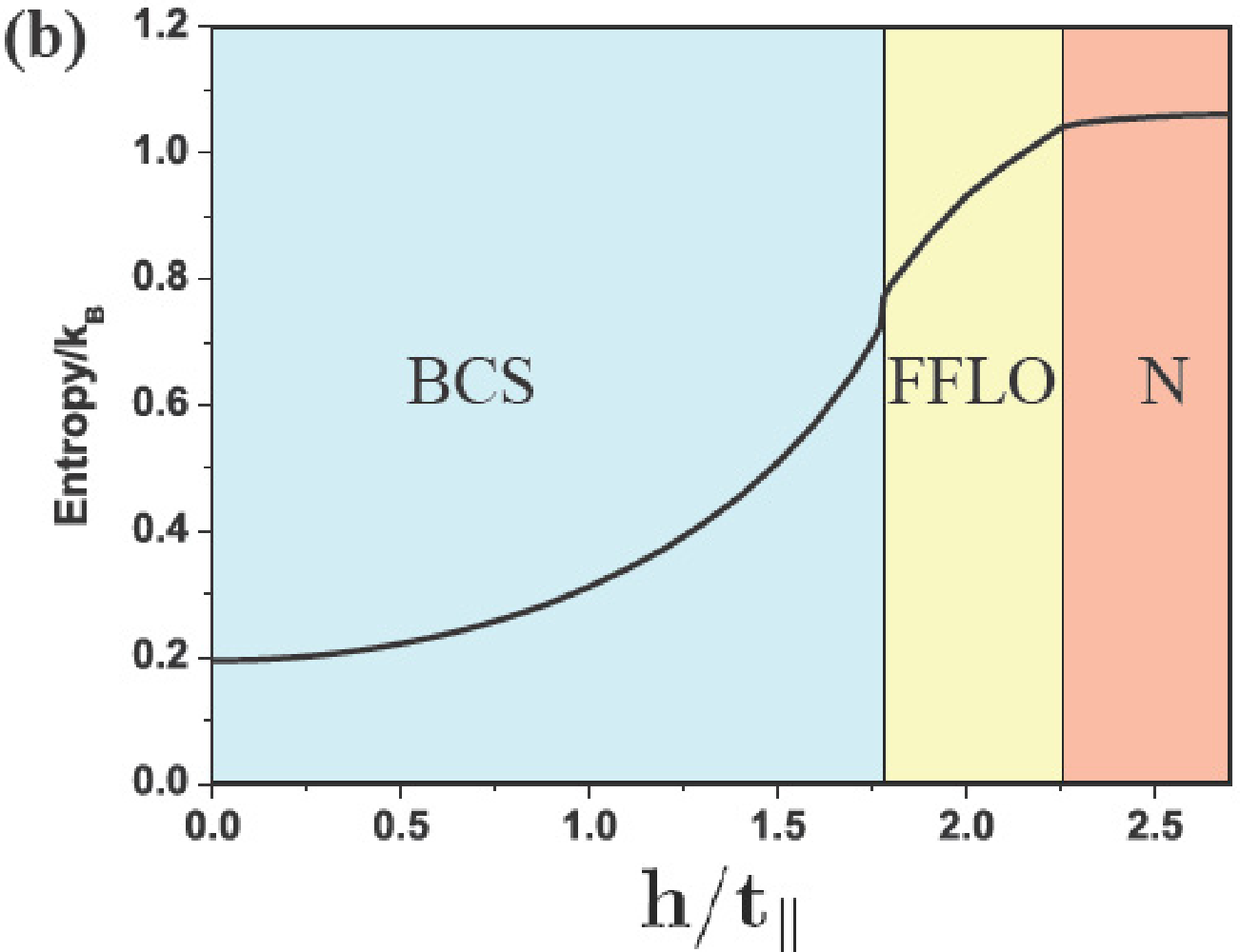}
\caption{(Color online) (a) The finite temperature phase digram for
the 3D $p$-orbital bands with $U/t_\pp=-2.4$, and $\mu=0$.
$T_c/t_\pp=0.84$ is the critical temperature for the BCS state. (b)
The entropy density $S/k_B$ v.s. $h/t_\pp$ at a finite temperature
of $T/t_\parallel=0.4$ with parameters $\mu=0$ and
$U/t_\parallel=-2.4$.}. \label{fig:finite_Temp}
\end{figure}

The physics of the FFLO states in the $p$-orbital bands in the 3D
cubic optical lattices is similar. The 3D $p$-orbital Hamiltonian is
similar to Eq. \ref{eq:bandsquare} and Eq. \ref{eq:porbital_int},
and further augmented by a new orbital $p_z$. The pair density
wavevectors are along the body diagonal directions, {\it i.e.}, the
$\pm[111]$ or other equivalent directions. Similar to the
Eq.(\ref{eq:delta}), the FFLO state in 3D cubic optical lattice is
characterized by the sinusoidal order parameter configuration as:
\begin{equation}
\Delta_x(\vec r)=\Delta_y(\vec r) =\Delta_z(\vec r)=|\Delta| \cos
(\vec Q \cdot \vec r). \label{eq:delta2}
\end{equation}
where $\vec Q=(\pm \delta k_f,\pm \delta k_f,\pm \delta k_f)$ are
along the body-diagonal directions.

In the 3D $p$-orbital bands, the long range ordered BCS and FFLO
states survive at finite temperatures, and  mean-field theory works
qualitatively well. We present the finite temperature phase diagram
of the competing orders at $U/t_\pp=-2.4$ and $\mu=0$ in Fig.
\ref{fig:finite_Temp} (a). The FFLO state can also survive to finite
critical temperatures at the same order of $T_c$. We further present
the entropy $S$ v.s. $h$ for different competing orders at a fixed
temperature $T/t_\parallel=0.4$ in Fig. \ref{fig:finite_Temp} (b).
The FFLO state has a large value of entropy density due to the extra
unpaired majority fermions, which interpolate between the BCS and
the fully polarized normal state. This greatly increases the
accessibility of the FFLO state in the cold atom optical lattices.
The transition between the BCS and the FFLO states is first order as
indicated by the discontinuity of entropy in Fig.
\ref{fig:finite_Temp} (b).

At last, we discuss experiment realizations and detections.
The $p$-band fermion systems can be realized by first preparing
enough number of atoms to fully fill the $s$-orbital band, thus the
extra particles will fill the $p$-bands.
The attractive interaction can be achieved through Feshbach
resonances in lattices \cite{liao2009,chinjk2006}, whose strength
can be tuned comparable to the band width of $4t_\parallel \approx 0.5E_R$
at $V_0/E_R\approx 15$ \cite{isacsson2005},  but still small compared
to band gaps which is around several $E_R$.
Our work predicts a large stable parameter regime for the FFLO
states.
These states can be detected by many methods
\cite{yang2005,mizushima2005,bakhtiari2008},
such as the direct imaging of the density profile oscillations of
each of the fermion components, the $rf$ spectroscopy measurement on
the collective modes, converting Cooper pairs into molecules and
measuring their momenta, the shot-noise correlation of the Fermi
momenta between $\vec k$ and $-\vec k \pm \vec Q$, etc. In
particular, the recent development of the {\it in situ} imaging
methods with the single site resolution \cite{gemelke2009,bakr2009}
can be used to accurately determine the spatial oscillation of the
FFLO states.

In summary, we have studied the competing orders among the FFLO,
the BCS, and the normal states in the spin imbalanced $p$-orbital
band systems in both 2D and 3D.
The FFLO states are stabilized by the combined effects of the quasi-1D
Fermi surfaces and the high dimensional phase coherence
built up by the inter-orbital interactions.
The pairing density wavevectors are along the diagonal directions
to facilitate the maximal inter-orbital pairing phase coherence.
The FFLO states are robust with many realistic experimental effects
including the confining trap, the small transverse $\pi$-bonding,
and finite temperatures.
It would be nice to realize the 2D and 3D stable FFLO phases in the
$p$-orbital bands in optical lattices which have not been identified
in solid state systems yet.

Z.C. and C. W. are supported by the Sloan Research Foundation,
NSF-DMR-03-42832 and AFOSR-YIP.
Y. P. W. is supported by NSFC and 973-project of MOST (China).


\end{document}